\RequirePackage{fix-cm}
\documentclass[twocolumn,epjc3]{svjour3} 
\usepackage{graphicx}
\usepackage{amssymb}
\usepackage{amsmath}
\usepackage{epstopdf}
\usepackage{array}
\usepackage{upgreek}
\newcolumntype{L}[1]{>{\raggedright\let\newline\\\arraybackslash\hspace{0pt}}m{#1}}
\newcolumntype{C}[1]{>{\centering\let\newline\\\arraybackslash\hspace{0pt}}m{#1}}
\newcolumntype{R}[1]{>{\raggedleft\let\newline\\\arraybackslash\hspace{0pt}}m{#1}}
\usepackage[table]{xcolor}
\usepackage{booktabs}
\usepackage{multirow}
\usepackage[colorlinks,citecolor=red,urlcolor=blue,bookmarks=false,hypertexnames=true]{hyperref} 
\usepackage{doi} 
\colorlet{tableheadcolor}{gray!25}
\colorlet{tablerowcolor}{gray!12.5}
\usepackage{tablefootnote}
\usepackage{siunitx}
\usepackage{lipsum}

\newcommand{\Nuc}[2]{\ensuremath{^{#2}\mbox{#1}}}

\journalname{Eur. Phys. J. C}
\title{The liquid-argon scintillation pulseshape in DEAP-3600}

\newcommand{\UofA}{Department of Physics, University of Alberta, Edmonton, Alberta, T6G 2R3, Canada}
\newcommand{\astrocent}{AstroCeNT, Nicolaus Copernicus Astronomical Center of the Polish Academy of Sciences, Rektorska 4, 00-614 Warsaw, Poland}
\newcommand{\CNL}{Canadian Nuclear Laboratories Ltd, Chalk River, Ontario, K0J 1J0, Canada}
\newcommand{\CIEMAT}{Centro de Investigaciones Energ\'eticas, Medioambientales y Tecnol\'ogicas, Madrid 28040, Spain}
\newcommand{\CU}{Department of Physics, Carleton University, Ottawa, Ontario, K1S 5B6, Canada}
\newcommand{\LNGSA}{INFN Laboratori Nazionali del Gran Sasso, Assergi (AQ) 67100, Italy}
\newcommand{\RHUL}{Royal Holloway University London, Egham Hill, Egham, Surrey TW20 0EX, United Kingdom}
\newcommand{\LU}{Department of Physics and Astronomy, Laurentian University, Sudbury, Ontario, P3E 2C6, Canada}
\newcommand{\UNAM}{Instituto de F\'isica, Universidad Nacional Aut\'onoma de M\'exico, A.\,P.~20-364, M\'exico D.\,F.~01000, Mexico}
\newcommand{\INFN}{INFN Napoli, Napoli 80126, Italy}
\newcommand{\PRISMA}{PRISMA Cluster of Excellence and Institut f\"ur Kernphysik, Johannes Gutenberg-Universit\"at Mainz, 55128 Mainz, Germany}
\newcommand{\PU}{Physics Department, Princeton University, Princeton, NJ 08544, USA}
\newcommand{\QU}{Department of Physics, Engineering Physics, and Astronomy, Queen's University, Kingston, Ontario, K7L 3N6, Canada}
\newcommand{\RAL}{Rutherford Appleton Laboratory, Harwell Oxford, Didcot OX11 0QX, United Kingdom}
\newcommand{\SL}{SNOLAB, Lively, Ontario, P3Y 1N2, Canada}
\newcommand{\Sussex}{University of Sussex, Sussex House, Brighton, East Sussex BN1 9RH, United Kingdom}
\newcommand{\TRIUMF}{TRIUMF, Vancouver, British Columbia, V6T 2A3, Canada}
\newcommand{\TUM}{Department of Physics, E15, Technische Universit\"at M\"unchen, 85748 Garching, Germany}
\newcommand{\Napoli}{Physics Department, Universit\`a degli Studi ``Federico II'' di Napoli, Napoli 80126, Italy}
\newcommand{\LBLNSD}{Nuclear Science Division, Lawrence Berkeley National Laboratory, Berkeley, CA 94720}
\newcommand{\kurchatov}{National Research Centre Kurchatov Institute, Moscow 123182, Russia}
\newcommand{\MEPhI}{National Research Nuclear University MEPhI, Moscow 115409, Russia}
\newcommand{\Cagliari}{INFN Cagliari, Cagliari 09042, Italy}
\begin{document}


\author{
The DEAP collaboration: P.~Adhikari\thanksref{CU} \and
R.~Ajaj\thanksref{CU} 
\and
G.\,R.~Araujo\thanksref{TUM}
\and
M.~Batygov\thanksref{LU} 
\and
B.~Beltran\thanksref{UofA} 
\and
C.\,E.~Bina\thanksref{UofA} \and
M.\,G.~Boulay\thanksref{CU,QU} \and
B.~Broerman\thanksref{QU} \and
J.\,F.~Bueno\thanksref{UofA} \and
A.~Butcher\thanksref{RHUL}
% ...
} 

\institute{\UofA \label{UofA} 
% ...
\and
\CU \label{CU}
% ...
\and
\LU \label{LU}
% ...
\and
\QU \label{QU} \and
\RHUL \label{RHUL}
% ...
\and
\TUM \label{TUM}
}
\thankstext[$\dagger$]{t1}{Deceased.}
\thankstext[*]{t2}{Currently at \Cagliari}
\thankstext[**]{t3}{Currently at \LBLNSD}
\maketitle


\begin{abstract}

DEAP-3600 is a liquid-argon scintillation detector looking for dark matter. Scintillation events in the liquid argon (LAr) are registered by 255 photomultiplier tubes (PMTs), and pulseshape discrimination (PSD) is used to suppress electromagnetic background events. The excellent PSD performance of LAr makes it a viable target for dark matter searches, and the LAr scintillation pulseshape discussed here is the basis of PSD.

The observed pulseshape is a combination of LAr scintillation physics with detector effects. We present a model for the pulseshape of electromagnetic background events in the energy region of interest for dark matter searches.
The model is composed of a) LAr scintillation physics, including the so-called intermediate component, b) the time response of the TPB wavelength shifter, including delayed TPB emission at $\mathcal O$(ms) time-scales, and c) PMT response.

TPB is the wavelength shifter of choice in most LAr detectors. We find that approximately 10\% of the intensity of the wavelength-shifted light is in a long-lived state of TPB. This causes light from an event to spill into subsequent events to an extent not usually accounted for in the design and data analysis of LAr-based detectors.

\end{abstract}

\section{Introduction} \label{sec:intro}


Several ongoing and planned particle physics experiments, in particular those looking for rare interactions, use liquid argon (LAr) as a particle detection medium~\cite{Fiorillo:2006bt,Gary:2013bj,Agostini2015jf,Meyers:2015dc,Aalseth:2017um,gerda,legend,Collaboration:2016ty,Akimov:2018ghi,Ajaj:2019wi}. Liquid argon is a bright scintillator that allows for excellent separation of electromagnetic interactions (`electron-recoils') from nuclear-recoil events even at low energies, based on differences in the scintillation pulseshape~\cite{d1psdpaper,Lippincott:2008uaa}. The pulseshape is the probability of photon detection as a function of time. Understanding the effects that influence features of the pulseshape helps with optimising the pulseshape discrimination (PSD) algorithm, and informs detector design and analysis choices.

The LAr pulseshape is well-known to have a double-exponential time structure originating from a short-lived singlet and a long-lived triplet state~\cite{Carvalho:1979tm,Kubota:1978kh,Morikawa:1989gv}. In addition, an intermediate component, which affects the pulseshape between approximately \SIrange{30}{100}{\nano\second}, is commonly observed~\cite{Hitachi:1983ja,Peiffer:2008zz,Lippincott:2008uaa,Acciarri:2010gm,Hofmann:2013hf}. Some authors attributed this component to late emission of the wavelength shifter 1,1,4,4-tetraphenyl-1,3-butadiene (TPB)~\cite{2015PhRvC..91c5503S}, making it an instrumental effect. However, the intermediate component was also observed in \cite{Hofmann:2013hf}, where the pulseshape was measured without the use of a wavelengh-shifter. This supports the hypothesis that the intermediate component is a feature intrinsic to LAr scintillation physics.

TPB absorbs the \SI{128}{\nano\meter} LAr scintillation photons and re-emits them at a peak wavelength of \SI{420}{\nano\meter}~\cite{Burton:1973up,Davies:1996vd}, where photon detection is easier. The TPB emission time is usually considered to be comparable to the LAr singlet decay time. TPB re-emission components at timescales much larger than $\mathcal{O}$(ns) (larger than the timescale of the intermediate component) were first reported for excitation with alpha particles~\cite{TPBpaper,Veloce:2015slj}, and more recently at $\mathcal{O}$(ms) timescales also for excitation with UV light~\cite{2015PhRvC..91c5503S,Stanford:2018un,Asaadi:2018vz}. Those measurements were done in dedicated small-scale setups. The intensity of this delayed TPB emission component is much smaller than that of the LAr triplet decay time close to the event peak, so that it is not a dominant effect in analysis. However, because it is so long lived, it causes light from one event to spill into subsequent events, which does result in a noticeable effect on for example the energy calibration.

This work corroborates the model from \cite{Hofmann:2013hf}, which attributes the intermediate component to a feature intrinsic to LAr, and confirms the existing evidence for delayed TPB emission. Both are measured for the first time here in a large LAr-based particle detector.

The pulseshapes contain information on the LAr excimer decay itself but also on detector properties. Once the contributions to the pulseshape are understood, it can be used to extract a) the LAr triplet lifetime, which serves as purity monitor for the LAr target, and b) the magnitude of instrumental effects, to monitor the stability of the light collection and detection system.

 We discuss the scintillation pulseshape from \Nuc{Ar}{39} beta decays, as measured in the DEAP-3600 single-phase LAr dark matter detector~\cite{detectorpaper}, starting at the time of the event peak out to \SI{160}{\micro\second}. We focus on overall effects dominating the pulseshape in different time windows, and disregard or simplify subdominant systematic effects to obtain the simplest model that describes the overall observed features well enough to inform analysis and simulation of DEAP data.

\section{The DEAP-3600 detector} \label{sec:experiment}

The DEAP-3600 detector is described in detail in \cite{detectorpaper}. We limit the description here to only the parts relevant to this work.

The centre of the DEAP-3600 detector is a spherical volume \SI{170}{cm} in diameter, which contains \SI{3.3}{tonnes} of LAr. The scintillation light created in the LAr travels through the argon volume until it reaches the surface of the acrylic vessel (AV) containing the argon. The inside acrylic surface is coated with a \SI{3}{\micro\meter} thick layer of the organic wavelength shifter TPB~\cite{Broerman:2017hf}.
The wavelength-shifted scintillation light is transmitted to the light detectors through a total of \SI{50}{cm} of acrylic in the form of the AV and acrylic light guides. The 255 cylindrical light guides protrude radially from the acrylic vessel.

A Hamamatsu R5912 high quantum efficiency photomultiplier tube (PMT) is optically coupled to the end of each light guide. The PMTs are shielded from magnetic fields by individual FINEMET$^\text{\textregistered}$\cite{finemet} collars, and by magnetic compensation coils located just outside the detector. Additional copper collars prevent large temperature gradients across the length of the PMTs. The PMTs operate at temperatures from \SIrange{-20}{5}{\celsius}.

\section{The pulseshape}\label{sec:averageps}
\Nuc{Ar}{39} is a $\beta$-emitter that occurs naturally in the atmospheric argon used in the DEAP-3600 detector. The \Nuc{Ar}{39} $\beta$ decays provide a high-statistics sample of LAr scintillation in response to electrons with energies between the trigger threshold of the detector and the \Nuc{Ar}{39} endpoint at \SI{565}{\kilo\electronvolt_{ee}} \cite{Ajaj:2019to}. We select events in the approximate energy window used for dark matter search, between \SI{13}{\kilo\electronvolt_{ee}} and \SI{40}{\kilo\electronvolt_{ee}} for this analysis. The unit \SI{}{\kilo\electronvolt_{ee}} refers to the energy scale for electromagnetic interactions. It is related to the energy scale for nuclear recoil events through the quenching factor \cite{Hitachi:1992ve}.

In DEAP-3600, a trigger is generated and data are collected when a total charge equal to the mean charge of approximately 19 photoelectrons is detected in a sliding \SI{177}{ns} window. Upon triggering on an event, the data acquisition system (DAQ) records the voltage on each PMT every \SI{4}{\nano\second}. The digitization is set such that the event peak occurs approximately \SI{2.5}{\micro\second} into the digitized PMT traces. Normal dark matter search data have a \SI{16}{\micro\second} long event window. For the analysis presented here, approximately \SI{36}{hours} of data were recorded with a \SI{200}{\micro\second} long window. The DAQ does not re-trigger within the digitization window of an event, even if the trigger condition is met again.

\begin{figure}[htbp]
\begin{center}
\includegraphics[width=\columnwidth]{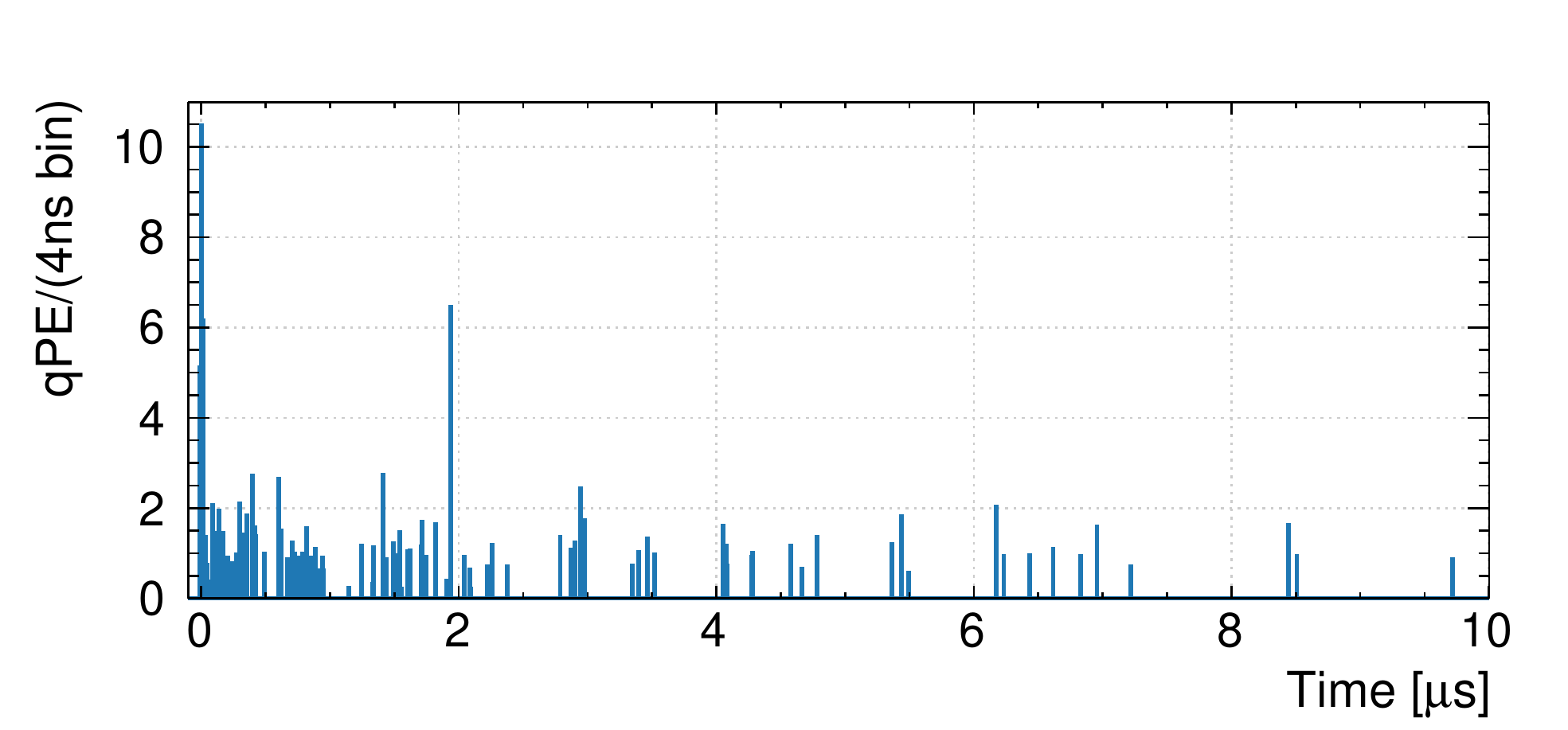}
\caption{A typical trace from an approximately \SI{20}{keV_{ee}} \Nuc{Ar}{39} event. The pulses from all PMTs are shown together.}
\label{fig:trace}
\end{center}
\end{figure}

A pulse-finding algorithm is applied to the digitized PMT traces to find the charge and time of each pulse. The pulse charge is converted to photoelectrons through division by the average single-photoelectron charge of the PMT. The resulting variable, called qPE, contains true photoelectrons, but also PMT dark noise and afterpulsing. The event peak time is determined based on the time when most qPE are detected. The qPE arrival times are then corrected such that the event peak occurs at t~=~\SI{0}{\nano\second}. An example for the resulting calibrated trace, from an electron-recoil event of approximately \SI{20}{keV_{ee}}, is shown in Fig.~\ref{fig:trace}. We construct the pulseshape by summing the calibrated traces from many events. The resulting curve is not normalized to one, but normalized such that units of rate or qPE per bin are obtained, since these quantities are more relevant here than photon detection probabilities. The bin width in some of the histograms shown in later sections is increased in regions of low intensity to reduce the uncertainty from counting statistics. The bin contents are then weighted by the bin width to obtain the correct unit again.

In this analysis, we consider events with a total number of qPE from \SIrange{100}{300}{qPE}. Only in this section, a pulseshape from events with \SIrange{500}{800}{qPE} is also shown. The PMT response is linear at these low numbers of qPE. Events additionally had to pass the following data quality cuts: i) low-level: e.g. stable baselines on all PMTs and success of pulse-finding algorithm, ii) reconstructed event position: inside the bulk of the LAr volume, far enough from the surface that no PMT sees more than 20\% of the total light in the event, iii) pile-up cuts: only a single event peak in the pulseshape, at most 3~photons detected in the first \SI{1.6}{\micro\second} of the trace (the event peak occurs \SI{2.6}{\micro\second} into the trace), an event time close to the DAQ trigger time, and at least \SI{20}{\micro\second} (for \SI{16}{\micro\second} long traces) or \SI{200}{\micro\second} (for \SI{200}{\micro\second} long traces) elapsed since the previous triggered event.

\begin{figure}[htbp]
\centering
\includegraphics[width=\columnwidth]{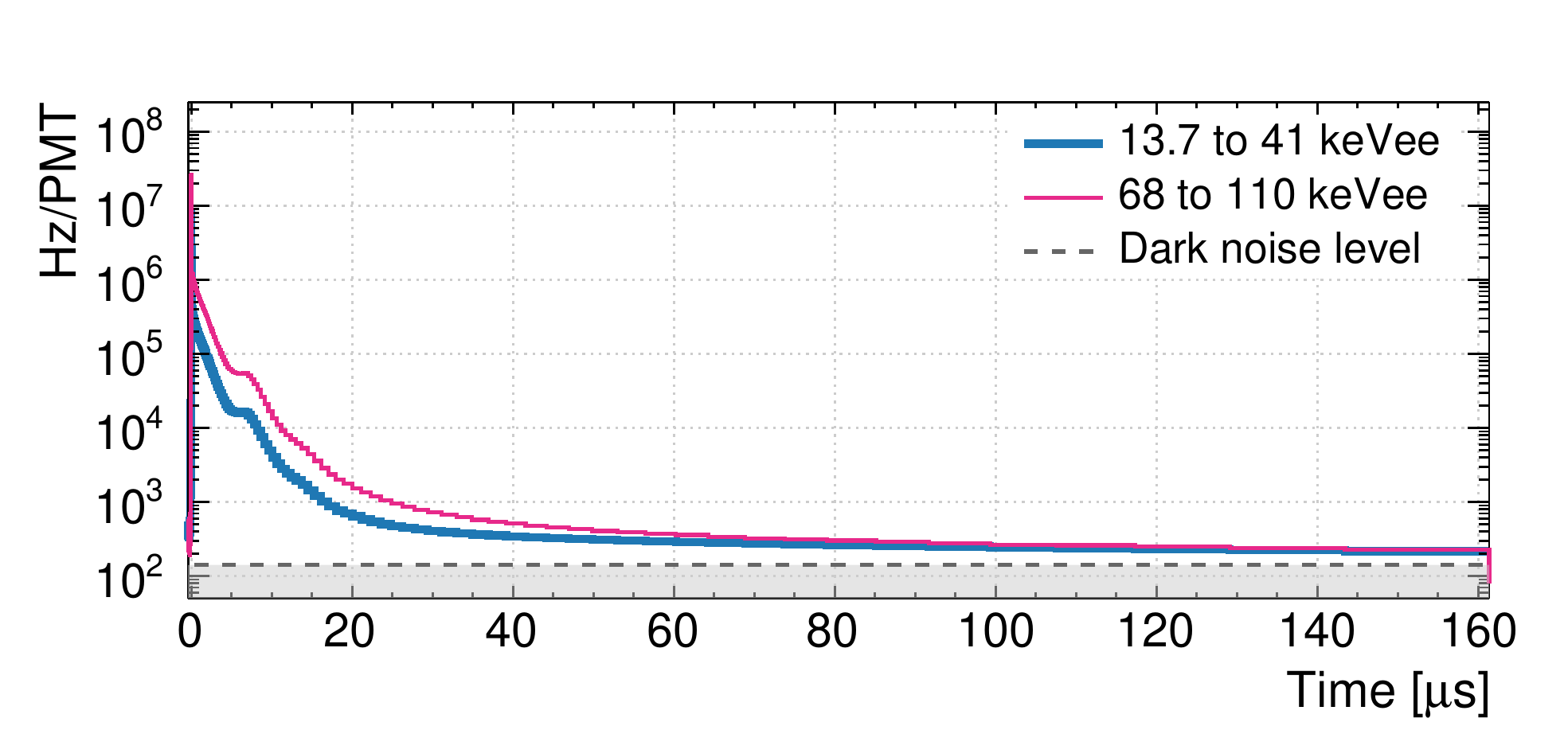}
\caption{The average pulseshapes in different energy windows, normalized to the number of events in each histogram and to the total number of PMTs. Approximately \SI{200000}{}~events were used for each energy window. The mean dark noise level of the PMTs is indicated by the dashed line. The pulseshapes do not reach the dark noise level even after \SI{160}{\micro\second} due to an additional component of uncorrelated light as will be discussed in Sect.~\ref{sec:fitmodel}.}
\label{fig:ps}
\end{figure}

Figure~\ref{fig:ps} shows the pulseshape in two energy windows. The histograms are normalized to show rate per PMT. The event peak at t~=~\SI{0}{\nano\second}, dominated by the LAr singlet and intermediate decay, is followed by the LAr triplet decay-dominated region up to approximately \SI{5}{\micro\second}. Features at approximately \SI{6.5}{\micro\second} and \SI{13}{\micro\second} are due to PMT afterpulsing. At $t\geq$\SI{14}{\micro\second}, the light intensity is still an order of magnitude above the PMT dark noise level, and scales with the event energy as expected for light correlated with the event.
Even \SI{160}{\micro\second} after the event peak, the light level has not subsided to the level of PMT dark noise, though the intensities from both energy windows approach the same level here. This indicates the presence of a source of noise in addition to uncorrelated PMT dark noise. The dark noise rate is taken from Fig.~\ref{fig:megatrace} and will be discussed together with the origin of the additional noise component in Sect.~\ref{subsec:verylateTPB}.

\section{Effects contributing to the pulseshape}\label{sec:fitmodel}
In this section, we describe the dominant effects that influence the observed pulseshape and provide mathematical descriptions for their time structures. 

\subsection{Liquid-argon scintillation}

We use the standard double-exponential model for the argon scintillation time structure, but add the empirical second term in Eq.~\ref{eq:purelar} to describe the intermediate component proposed in \cite{Hofmann:2013hf}. This component is modified only to normalize the function but we otherwise follow their nomenclature. The time structure of the pure LAr scintillation signal is then:

\begin{align}
I_{\text{LAr}}(t) =  \frac{R_s}{\tau_s} e^{-t/\tau_s} + &\frac{1-R_s-R_{t}}{(1 + t/\tau_{rec})^2}\frac{1}{\tau_{rec}} \nonumber \\
         &+ \frac{R_{t}}{\tau_t} e^{-t/\tau_t} \label{eq:purelar}\; ,
\end{align}
where $\tau_s$ and $\tau_t$ are the LAr singlet and triplet lifetimes.  $R_{s,t}$ are the relative intensity of each component. In \cite{Hofmann:2013hf}, the intermediate component is attributed to electrons that were ejected out of the immediate reach of their ions' attractive electric fields, and re-combine only after a random walk. $\tau_{rec}$ is the characteristic time for this recombination process. The term for the intermediate component is set to 0 for times later than \SI{1.2}{\micro\second} because it is numerically insignificant for larger times.

The work of~\cite{Hofmann:2013hf} is based on ~\cite{Ribitzki:1994de,Kubota:1979gk,hofmann} in which the following four assumptions are laid out. We quote these from~\cite{Ribitzki:1994de} verbatim:
\begin{enumerate}
    \item The electrons have cooled down to room temperature at the very end of the collisional processes in the target gas.
    \item  The electrons are homogeneously distributed in the
observed volume.
    \item The electron density is equal to the density of
molecular ions.
    \item The time scale for photon emission is dominated by
dissociative recombination.
\end{enumerate}

\subsection{TPB fluorescence}
LAr scintillation photons are absorbed by the TPB and re-emitted in the visible spectral region. The lifetimes of the prompt TPB emission and of the LAr singlet decays are both at the order of a few ns and cannot be separately resolved here. We therefore consider the prompt TPB emission a delta function. This changes the interpreation of the singlet lifetime from Eq.~\ref{eq:purelar} as will be discussion in Sect.~\ref{sec:fullmodel}. We use the model from \cite{Stanford:2018un} for the time structure of the delayed TPB emission:

\begin{align}
I_{\text{TPB}}&(t) =  (1 - R_\text{TPB})\delta (t) \nonumber \\ 
 & +\frac{ R_\text{TPB}\cdot N_\text{TPB} \cdot e^{-2t/\tau_T} }{ { 1 + A_{\text{TPB}}[ Ei(-\frac{t+t_a}{\tau_T} - Ei(-\frac{t_a}{\tau_T}) ] }^2 (1 + t/t_a) }  \label{eq:TPBPrinceton} \; .
\end{align}

where $N_\text{TPB}$ is a normalization to make the integral of $I_{\text{TPB}}(t)$ equal to 1, $R_\text{TPB}$ is the probability that the photon will be re-emitted late, and $Ei$ is the exponential integral. We refer the reader to \cite{Stanford:2018un} for more detailed explanation of the terms in the equation.

\subsection{Detector geometry and PMT noise}
The geometry of the DEAP-3600 detector results in a characteristic photon time distribution due to scattering~\cite{pmtpaper}, with the intensity of observed photons dropping to 10\% of the maximum within approximately \SI{15}{\nano\second}. 
Once a photon hits a PMT, the signal from the resulting photoelectron can be delayed when photoelectrons recoil on a dynode instead of, or in addition to, releasing secondary electrons. The resulting double and late pulsing in the PMTs causes an approximately gaussian peak centered at \SI{58}{\nano\second} after the nominal arrival time. 

The time structures from scattering and double/late pulsing are further smeared with the uncertainty in the event peak time. The ability of the pulse finder to separate pulses that are close in time also affects the pulseshape somewhat.

Photon scattering, early pulsing, and late/double pulsing all occur at the same prompt time scale of approximately $\pm$\SI{50}{\nano\second}, so that we cannot make a precise measurement of any of the individual contributions. The goal of describing the peak structure mathematically is to obtain a function that can be used to estimate the total light intensity in the prompt region, and separate this contribution from effects with longer time constants.

The effective model for the prompt time structure consists of the sum of two gaussians:

\begin{align}
I_{\text{geo}}(t) =  \nu_\text{DET}\cdot \text{Gaus}&(t, \mu_{\text{DET}},\sigma_{\text{DET}})\label{eq:igeo} \\ 
                    & + \nu_\text{DP}\cdot \text{Gaus}(t, \mu_\text{DP}, \sigma_\text{DP}) \nonumber \; ,
\end{align}

with $\nu_\text{DP} = 1 - \nu_\text{DET}$ and where $\nu_\text{DP}$ is the probability for a pulse to arrive late, and $\nu_\text{DET}$ in turn is the probability for a pulse to arrive at the nominal time.

Additionally, photoelectrons can skip a dynode in the PMT, leading to pulses that arrive early. This situation will be treated seperatedly later. 

The PMTs also produce correlated noise, so-called afterpulsing (AP). AP in the DEAP-3600 PMTs occurs in three broad time regions centered at approximately  \SI{0.5}{\micro\second}, \SI{1.7}{\micro\second}, and \SI{6.3}{\micro\second}.  In the calibration of the PMTs, each of these regions is modelled using a gaussian distribution~\cite{pmtpaper}. This simple model neglects small sub-structures within each AP region that are not relevant in analysis of single events, but become visible when looking at the summed pulseshape from many events. Nevertheless, we use the same model employed for PMT calibration here:

\begin{equation}
I_{\text{AP}}(t)   =  \sum_{i=1}^{3}\nu_{\text{AP}i} \cdot   \text{Gaus}(t, \mu_{\text{AP}i}, \sigma_{\text{AP}i})\label{eq:ap}
\end{equation}

where $i$ indicates the AP region, $\nu_\text{AP}$ is the probability for an AP to occur in the respective region, $\mu_\text{AP}$ is the time where the distribution is centered at, and $\sigma_\text{AP}$ is its width. 

We further consider AP of AP as the convolution of the AP distribution with itself:

\begin{align}
I_{\text{APofAP}}(t) & = I_{\text{AP}}(t) \otimes I_{\text{AP}}(t)
\end{align}

AP of AP of AP is numerically insignificant and therefore not considered.

The random PMT noise (dark noise, DN) is modeled as a single constant term:

\begin{equation}
I_{\text{DN}}(t) = r_{\text{DN}} \label{eq:dn}
\end{equation}

This term contains the constant rate of qPE from sources not correlated with the event that triggered the detector; this includes the true thermionic PMT dark noise, the light level from radioactive decays in the PMT glass and surrounding material (causing for example low level Cherenkov light in the acrylic light guides), light from LAr events at such low energies that they do not trigger the detector and are not removed by pile-up cuts, and AP from all these effects. 

\subsection{Very late correlated light from previous events}\label{subsec:verylateTPB}
Figure~\ref{fig:ps} shows that correlated light from \Nuc{Ar}{39} beta decay events is seen more than \SI{18}{\micro\second} after the event peak. Both the LAr triplet decay and PMT afterpulsing are well below dark noise level this late in the pulseshape. The observation is, however, what one expects if TPB has a very long-lived emission component: Each event selected in the energy windows discussed here is preceded by events that on average have a higher or much higher energy. The late TPB emission from these events will leak into following events, creating an average level of uncorrelated noise that is a function of the time since the previous event.

We use the term \emph{stray light} to denote uncorrelated noise that includes both dark noise and the average residual light level from previous events. The stray light level is a function of the time that passed since the previous event. To measure the stray light level, we make use of the fact that each event's trace starts \SI{2.6}{\micro\second} before the event peak. This pre-event window contains some of the light from previous events.  We group all events by the time that passed since the previous event, $\Delta t$. For each $\Delta t$, we then determine the total number of photons detected in the pre-event window over all those events, $N_{p}(\Delta t)$. The number of events in each group, $N_{ev}(\Delta t)$ is also recorded. This allows us to map the stray light level, in average number of photons detected, as a function of the time since the previous event, $I_\text{stray}(\Delta t)$, as 

\begin{equation}
I_\text{stray} (\Delta t) = \frac{ N_p(\Delta t)}{ N_{ev}(\Delta t)}
\end{equation}

In practice, a pre-event window of  \SIrange{-1.6}{-1.0}{\micro\second} is used, since the \SIrange{-2.6}{-1.6}{\micro\second} region is used in one of the pile-up cuts; using an overlapping window would bias the measurement.

\begin{figure}[htbp]
\begin{center}
\includegraphics[width=\columnwidth]{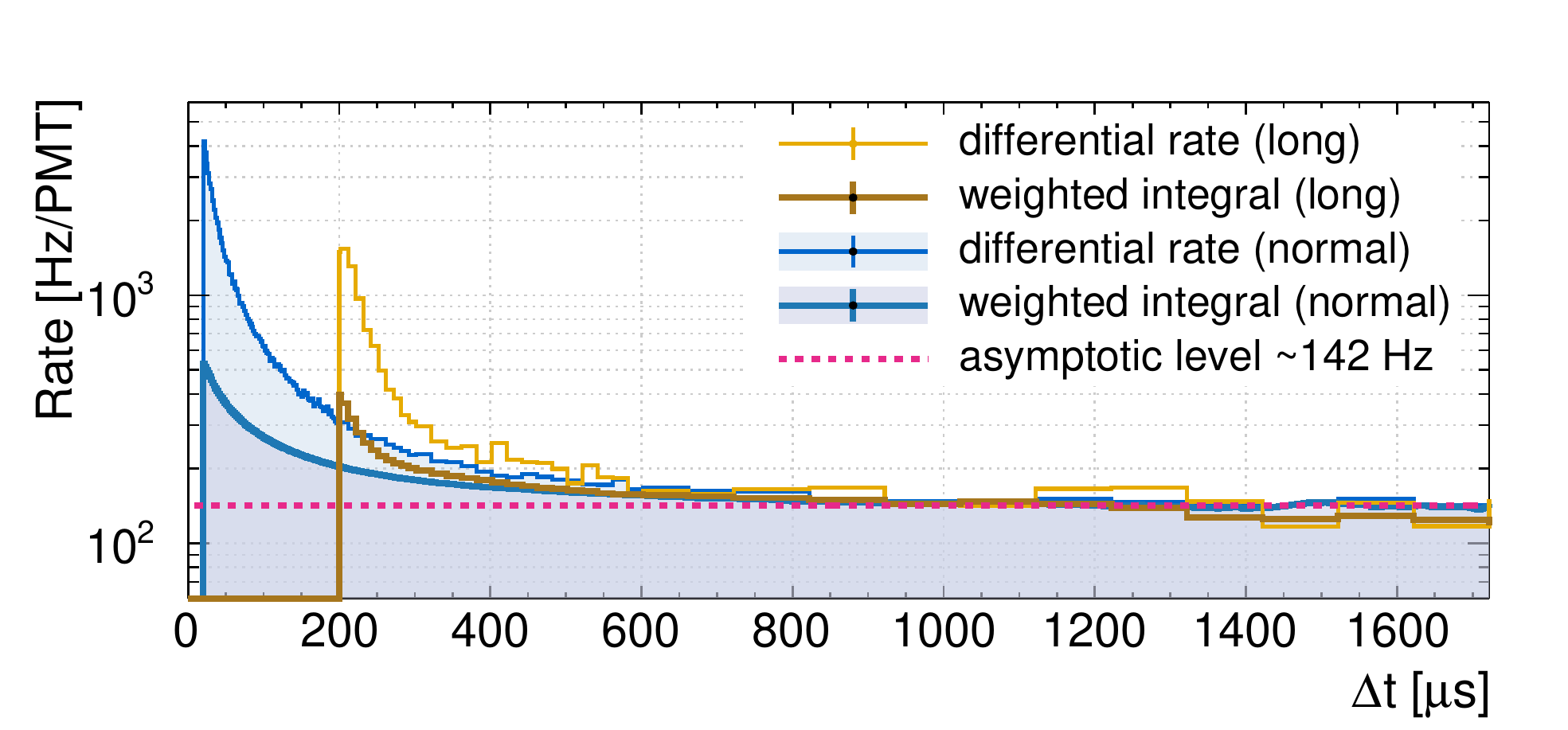}
\caption{The pre-event light level as a function of $\Delta t$, for normal (\SI{16}{\micro\second}) and long (\SI{200}{\micro\second}) PMT trace data. Both the differential rate, and the average light level above the $\Delta t$ value (see text) are shown. In all cases, the curves approach a light level of approximately \SI{142}{\hertz}. The peaks at \SI{20}{\micro\second} and \SI{200}{\micro\second} are due to pile-up (see text).}
\label{fig:megatrace}
\end{center}
\end{figure}

The result is converted to Hertz per PMT by dividing by the length of the sampling time window (\SI{0.6}{\micro\second}) and the number of PMTs. Figure~\ref{fig:megatrace} shows this differential pre-event light rate for events of \SI{200}{\micro\second} digitization window, as well as for normal detector data recorded with a \SI{16}{\micro\second} digitization window.

When we make the average pulseshapes as shown in Fig.~\ref{fig:ps} and in the figures in Sect.~\ref{sec:fitresult}, we accept all events with $\Delta t \geqslant\Delta t_{cut}$, where $\Delta t_{cut}$ is either \SI{20}{\micro\second} or \SI{200}{\micro\second}, depending on the dataset. So we need the stray light level in an event when the previous event occurred \emph{at least} $\Delta t$ before. This is obtained by determining the average stray light level above a given value of $\Delta t$:

\begin{equation}
\overline{I}_\text{stray} (\Delta t) = \frac{ \int_{\Delta t}^{\infty} N_p(\Delta t') d\Delta t'}{ \int_{\Delta t}^{\infty} N_{ev}(\Delta t')d\Delta t' }
\end{equation}
This distribution is called the \emph{weighted integral} in Fig.~\ref{fig:megatrace}, because $N_p(\Delta t)$ is implicitly weighted by the number of events at each $\Delta t$.

Finally, we assume that the pre-event light level obtained from events with $\Delta t\geqslant$\SI{21.6}{\micro\second} measures the stray light level at t=\SI{0}{ns} in events with $\Delta t\geqslant$\SI{20}{\micro\second}, and so on throughout the pulseshape. If $t$ is the time since the start of the event, that is the x-axis from Fig.~\ref{fig:ps}, and $\Delta t$ is the time axis of Fig.~\ref{fig:megatrace}, then the level of uncorrelated light at a given time $t$ in the event is $\overline{I}_\text{stray} (\Delta t = \Delta t_\text{cut} + 1.6\mu s + t)$.

The stray light level is highest near $\Delta t_\text{cut}$ due to pile-up in the preceding event. If $\Delta t$ was a perfectly accurate measure of the time difference to the last event, then we would not expect such a pronounced peak, and the curve for the long-digitization-window data would coincide with the curve for the normal data starting at $\Delta t$=\SI{200}{\micro\second}. However, $\Delta t$ is calculated to the last trigger, and the DAQ does not re-trigger within the digitization window. Hence we have to differentiate between an \emph{event}, that is an interaction that happens in the LAr and causes light emission, and a \emph{triggered event}, that is an interaction that also causes the DAQ to trigger PMT read-out. If an event occurs within another event's digitization window, the $\Delta t$ between triggers is larger than the actual time since the last event. The real $\Delta t$ can be as low as the time span that is the difference between $\Delta t_\text{cut}$ and the digitization window length. Since such a pile-up probability is constant in time, the uncorrelated light rate rises as the $\Delta t$ cut used approaches the length of the digitization window, regardless of the length of this window. This interpretation is corroborated by two observations: 1) the level this feature rises to is strongly influenced by the pile-up cut that removes events with too much light early in the trace, and 2) a toy Monte Carlo simulation that includes pile-up reproduces the shape and intensity of the feature. We also note that AP cannot cause the feature seen in Fig.~\ref{fig:megatrace} as it occurs at shorter time scales.

The intensity to which the pile-up feature rises is lower for the data taken with a \SI{200}{\micro\second} digitization window because the total intensity is the sum of the intensity from pile-up and the intensity of the delayed TPB emission (from the event that triggered the DAQ). The latter is smaller after \SI{200}{\micro\second} than it is after \SI{20}{\micro\second}.

\begin{figure}[htbp]
\begin{center}
\includegraphics[width=\columnwidth]{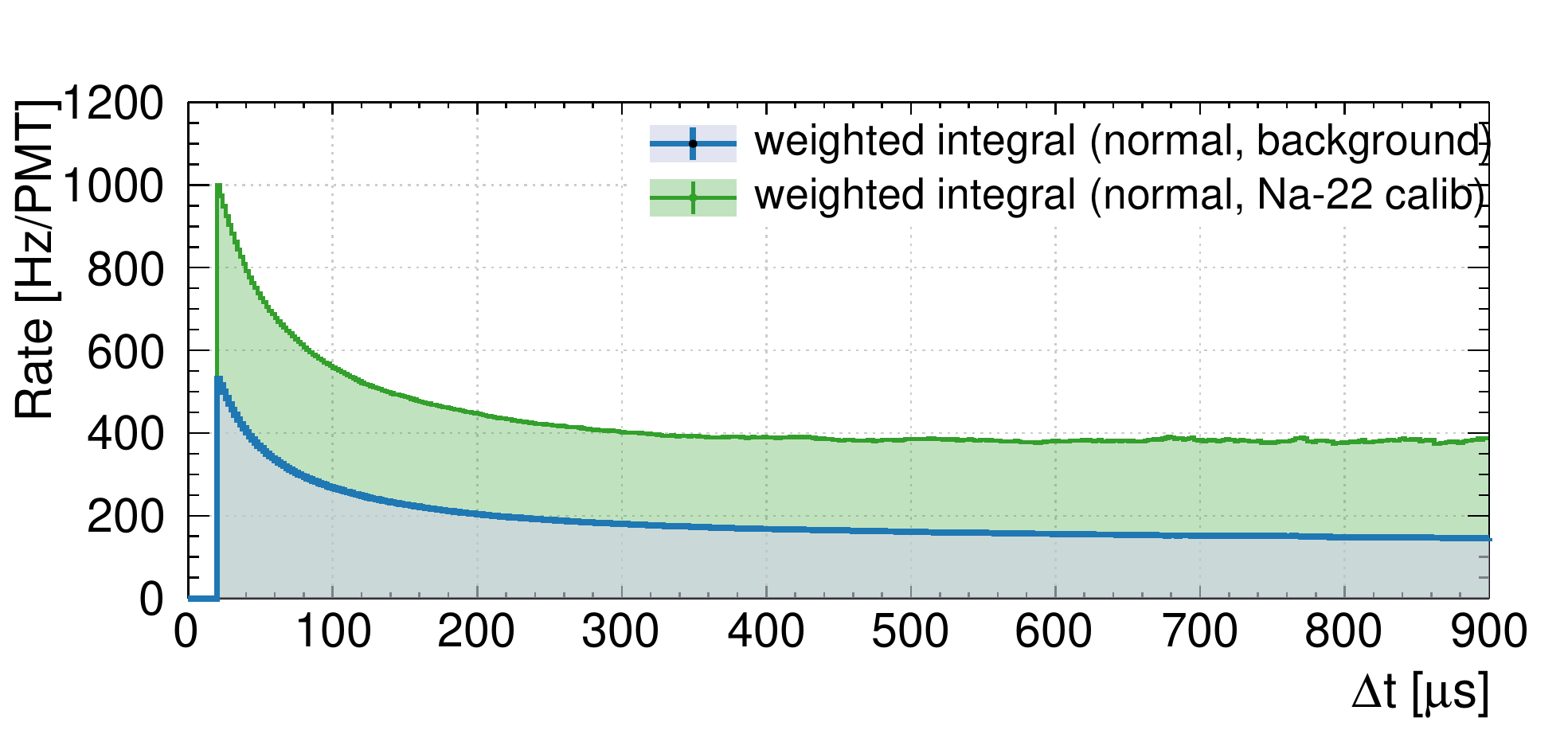}
\caption{The pre-event light level as a function of the $\Delta t$ cut, for a normal physics run (this is the same curve as in Fig.~\ref{fig:megatrace}) and data taken with a \Nuc{Na}{22} gamma calibration source. The level of stray light increases due to the increased total event rate.}
\label{fig:megatracena22}
\end{center}
\end{figure}

The pre-event pulse rate approaches the flat dark noise level at large values of $\Delta t$, i.e. it approaches $r_{\text{DN}}$ from Eq.~\eqref{eq:dn}. We use the $\overline{I}_\text{stray}$ histograms to describe the time structure of all uncorrelated light and thus do not need $r_{\text{DN}}$ in the fit model.

The curves in Fig.~\ref{fig:megatrace} change with event rate and spectrum. To illustrate this, Fig.~\ref{fig:megatracena22} shows a comparison between the stray light levels for normal physics data in a physics run (where the \Nuc{Ar}{39} provides the vast majority of events) to a run taken with a \Nuc{Na}{22} gamma calibration source. As expected, the rate of stray light increases, and it increases more strongly for values of $\Delta t$ near $\Delta t_\text{cut}$. Note that the source also induces an additional contribution to the flat dark noise level due to particles scattering on detector materials. Such scatters can cause Cherenkov photon emission, and reduce the energy of the particles as they reach the liquid argon, creating events with energy below the trigger threshold.

\subsection{Full model} \label{sec:fullmodel}

We describe the observed pulseshape by the convolution of detector effects with the LAr time structure. Detector effects in the prompt time region (\SIrange{-50}{100}{\nano\second}) are strongly degenerate in the fit. Therefore,  we replace the decay parameter of the singlet component in the LAr PDF ($\tau_s$) with a generalized decay time $\tau_p$, which stands in for all the effects with exponentially falling time-structures at the \SI{}{\nano\second} scale.

\begin{align}
I_{\text{PS}}(t) = &\eta\cdot\overline{I}_\text{stray}(\Delta T_{cut} + 1.6\mu s + t) \label{eq:ipsnoep} \\ 
   &+ \mathbf{I_0}\cdot \Big( I_{\text{LAr}}(t) \otimes I_{\text{TPB}}(t) \otimes I_{\text{geo}}(t)  \nonumber \\
   &\; + I_{\text{LAr}}(t) \otimes [I_{\text{AP}}(t) + I_{\text{APofAP}}(t)]\Big) \nonumber
\end{align}
where $\eta$ converts from Hz/PMT to pulse count.

AP following prompt photons creates a distinct peak in the pulseshape. AP in response to the LAr triplet decay is washed out but still creates a visible structure in the pulseshape. The TPB time structure is so extended that AP in response to it is washed out to the point where it is not visible in the pulseshape. Hence, AP in response to TPB delayed emission is not considered separately and the AP rate is absorbed in the overall TPB late emission probability.

A component due to early pulsing of the PMTs is added afterwards as $I_\text{EP}$. This component consists of the function $I_{\text{PS}}(t)$, shifted earlier in time and widened, since the early-pulsing has an intrinsic width. We model this by

\begin{align}
  I_\text{EP}&(t) = \mathbf{I_0}\cdot R_\text{EP}\\ \nonumber
  \cdot&(I_{\text{LAr}}(t-t_\text{EP}) \otimes I_{\text{TPB}}(t-t_\text{EP}) \otimes I'_{\text{geo}}(t-t_\text{EP}))  
  \label{eq:ipswep}
\end{align}

where in $I'_{\text{geo}}(t)$ the resolution of the gaussian is increased.

This component is not part of the fit, but is included when drawing the function:

\begin{align}
I'_{\text{PS}}(t) = I_\text{EP}(t) + I_{\text{PS}}(t)
\end{align}

In practice, all terms contributing less than approximately 0.5\% of the intensity at a given time are neglected in the evaluation of $I_{\text{PS}}$.

The model is constructed such that the total intensity $\mathbf{I_0}$ is the only parameter that determines the overall amplitude. The intensity of all individual components is relative to this intensity. Since the AP probability in DEAP-3600 PMTs relatively large (approximately 8\%), we re-calculate the intensities of the individual components after removing the AP contribution to the total intensity.

\section{Pulseshape fits}\label{sec:fitresult}

We consider the pulseshape from events in the energy region of interest for WIMP search. The pulseshape has up to $10^7$ qPE per bin. With this many counts, the standard statistical uncertainty of the square root of the number of counts is dwarfed compared to systematic effects as small as 0.03\% of the intensity in a bin. Since such small effects are not relevant when extracting information from the pulseshape or when simulating the detector response, they are not part of the model. Since the reduced $\chi^2$ is not a good indicator of goodness of fit in a situation where systematic errors dominate, we use the relative difference between the model and data instead of residuals to indicate how closely the model function describes the data. This quantity is shown below the figures in this section. The fit routine still attempts to minimize $\chi^2$; to improve convergence, the poissonian uncertainty in each bin is multiplied by a factor of $2$, forcing $\chi^2$ to be smaller than it would be with standard uncertainties. The choice of multiplication factor has no significant effect on the extracted parameters. Due to $\chi^2$ not being a good statistical measure, parameter uncertainties from the fit will not be correct, and are therefore not quoted.

Many of the effects that influence the pulseshape are correlated, therefore it is not possible to obtain best fit values with high confidence for all parameters in the model. The goal of the fit is rather to obtain parameters such that the model describes the pulseshape well enough to be useful.

The parameters of the delayed TPB emission (Eq.~\eqref{eq:TPBPrinceton}) are highly correlated with the LAr triplet decay time and with the AP rate. Therefore, we fix the TPB emission parameter values to those from \cite{Stanford:2018un}\footnote{The fit parameters changed significantly between the arXiv version and the published version of \cite{Stanford:2018un}. The original arXiv version fit the TPB pulseshape from UV excitation fairly well out to \SI{1}{\milli\second} and included an intermediate term that captured some of the residual LAr intermediate component we use in this paper. The published version of the paper focuses the fit on earlier times and removes the dedicated intermediate component. It no longer fits the later part of the UV-light induced TPB pulseshape very well, which is the part of relevance for this work. Therefore, after communication with the authors of \cite{Stanford:2018un}, the parameters we use here are those from the original arXiv version.} and only vary the total intensity of this component in the fit.

The AP rates and time structure (Eq.~\eqref{eq:ap}) were calibrated in-situ before the DEAP-3600 detector was filled with LAr. However, AP rates can change with time and with PMT temperature. Two AP distributions at times of approximately \SI{0.5}{\micro\second} and \SI{1.7}{\micro\second} have a small probability and thus only a small effect on the pulseshapes. Their parameters are fixed by the calibration. The AP distribution at approximately \SI{6.6}{\micro\second} dominates the pulseshape near that time, and the three parameters that describe it are varied in the fit to account for possible changes since the calibration.

The LAr triplet lifetime and the prompt lifetime (which accounts for the LAr and TPB prompt decay times, as well as the time structure from photons scattering in the detector) are varied in the fit. The recombination time of the intermediate component is not constrained to the times quoted in \cite{Hofmann:2013hf}, since the pulseshapes fit there are from interactions with protons or heavier nuclei, and we expect the shape to be different for low-energy electron-recoil events. The parameters that describe the prompt peak (Eq.~\ref{eq:igeo}) are all varied in the fit. 

The shape of the stray light intensity is taken from Fig.~\ref{fig:megatrace} and $\eta$ from Eq.~\ref{eq:ipsnoep} is adjusted such that the curves match the intensity of the pulseshape from \SIrange{-450}{-200}{\nano\second}.

The fit is done in several stages, where parameters that dominate either the prompt (\SIrange{0}{0.5}{\micro\second}), the intermediate (\SIrange{0.5}{8}{\micro\second}), or the late ($\geq$ \SI{8}{\micro\second}) region of the pulseshape are varied while all other parameters but the overall intensity and the singlet-to-triplet ratio are fixed in the fit. The set of parameters fit for one region is then fixed to its fit value when fitting the parameters for the next region. The prompt, intermediate, and late parameters are fit in turn and updated until the parameter values no longer change significantly. The early-pulsing component is added by manually adjusting the time difference, width, and early-pulsing probability to match the data at times before the peak. The prompt, intermediate, and late parameters were fit once more after adding this component.

\begin{figure}[htbp]
\begin{center}
\includegraphics[width=\columnwidth]{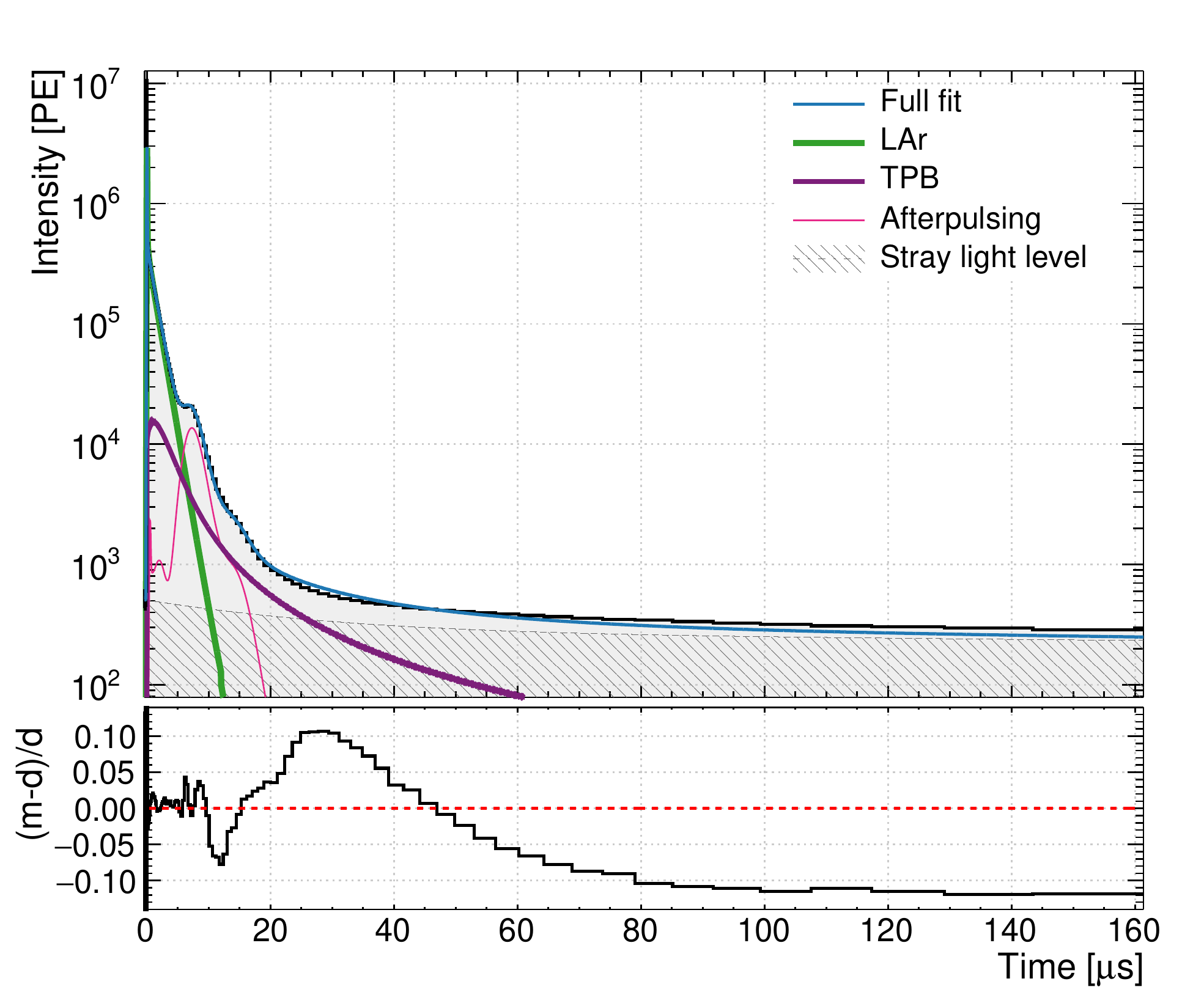}
\caption{At time scales beyond \SI{15}{\micro\second}, the pulseshape is dominated by the delayed TPB emission. At approximately \SI{30}{\micro\second}, the intensity of TPB emission has declined to the point where it is equal to the intensity of left-over late light from previous events.}
\label{fig:APSLongfit}
\end{center}
\end{figure}

\begin{figure}[htbp]
\begin{center}
\includegraphics[width=\columnwidth]{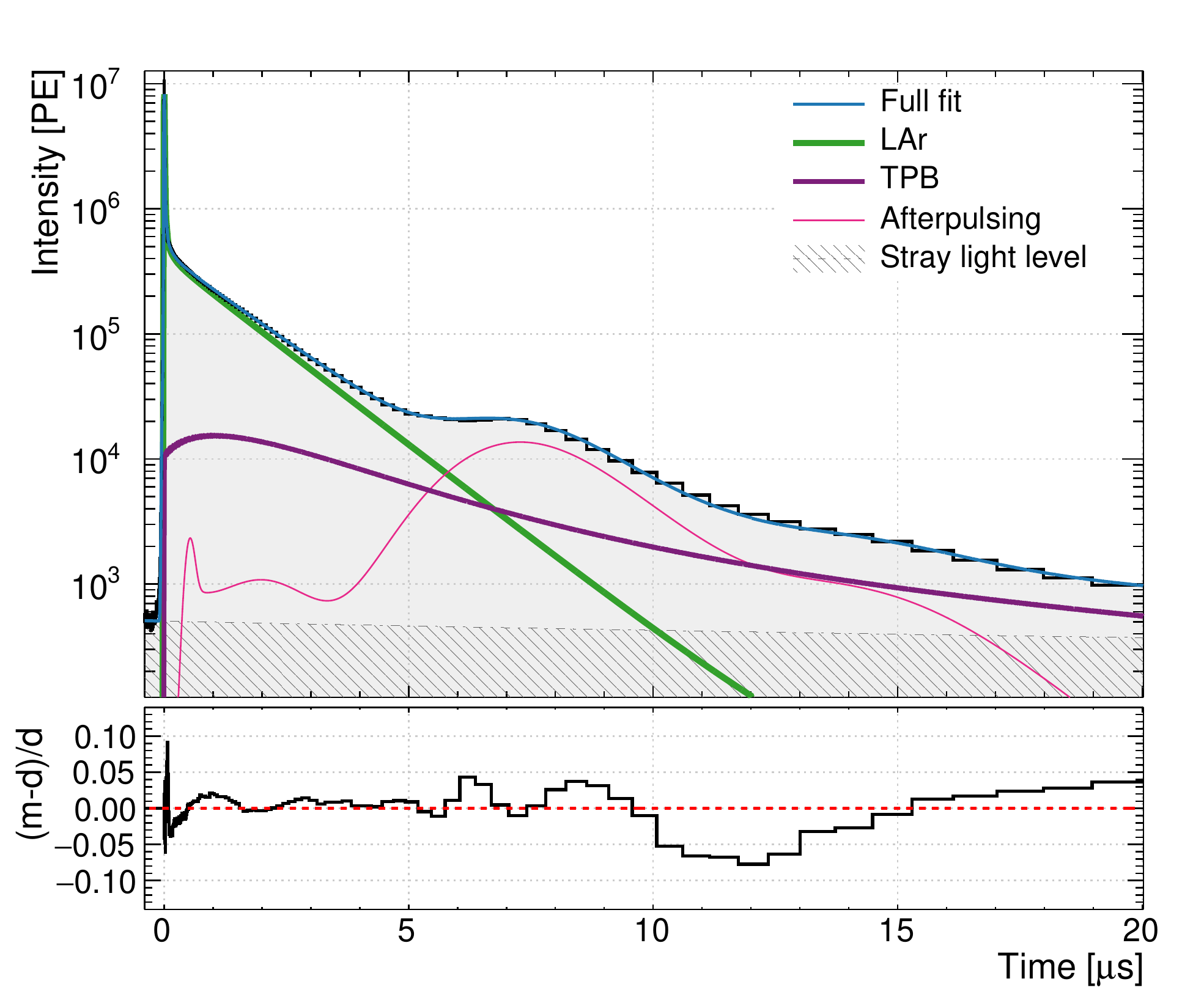}
\caption{From \SIrange{0.2}{5}{\micro\second}, the pulseshape is dominated by the LAr scintillation light. The region from \SIrange{5}{10}{\micro\second} is dominated by PMT afterpulsing. Starting at approximately \SI{13}{\micro\second}, the TPB delayed emission becomes significant. While the total event length in standard DEAP data is \SI{16}{\micro\second}, the analysis window on which PSD as well as event energy and position reconstruction is based is \SIrange{-0.03}{10}{\micro\second} with respect to the event peak.}
\label{fig:APSStandard}
\end{center}
\end{figure}

\begin{figure}[htbp]
\begin{center}
\includegraphics[width=\columnwidth]{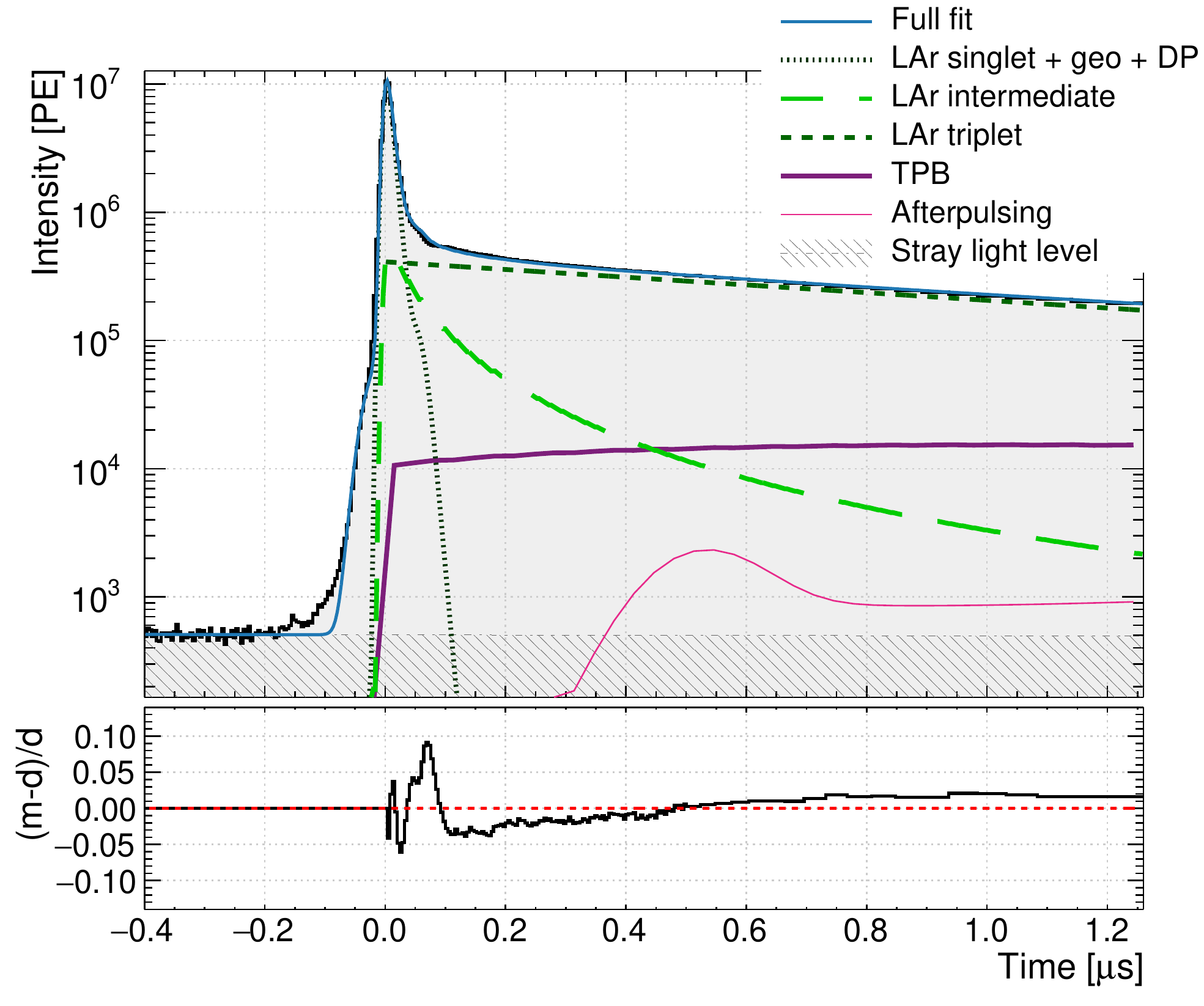}
\caption{The prompt region of the pulseshape, including the so-called intermediate times (approx. \SIrange{40}{100}{\nano\second}), is well described by our LAr scintillation model. The time region before \SI{-8}{\nano\second} is not part of the fit.}
\label{fig:APSshortfit}
\end{center}
\end{figure}

The full fit region is \SIrange{-0.008}{160}{\micro\second}. The initial estimates and the fit-out values for all model parameters are listed in Tab.~\ref{tab:parphysics} and \ref{tab:parinstrument}, and a comparison between model and data is shown at three different time ranges in Figs.~\ref{fig:APSLongfit} through \ref{fig:APSshortfit}.

\begin{table}[htp]
\caption{Start and fit parameters, LAr and TPB. Parameter uncertainties are not given, as explained in the text.}
\begin{center}
\begin{tabular}{|lll|lll|}
\multicolumn{3}{c}{LAr} & \multicolumn{3}{c}{TPB} \\
\hline
par & start & fit & par & start & fit \\
\hline
$R_p$ & 0.3 & 0.23 & $R_{TPB}$ & 0.06 & 0.1 \\
$\tau_p$ & \SI{3}{\nano\second} & \SI{8.2}{\nano\second} & $\tau_T$ & \SI{20e4}{\micro\second} & -- \\
$\tau_{rec}$ &  - & \SI{75.5}{\nano\second} & $t_a$ & \SI{12}{\micro\second} & -- \\
$R_t$ & 0.7 & 0.71 & $A_{TPB}$ & 4.6 & -- \\
$\tau_t$  & \SI{1564}{\nano\second} & \SI{1445}{\nano\second} &$\;$  & $\;$ & \\
\hline
\end{tabular}
\end{center}
\label{tab:parphysics}
\end{table}%

\begin{table}[htp]
\caption{Start and fit parameters, instrumental effects. Parameter uncertainties are not given, as explained in the text.}
\begin{center}
\begin{tabular}{|lll|lll|}
\multicolumn{3}{c}{AP} & \multicolumn{3}{c}{Detector} \\
\hline
par & start & fit & par & start & fit \\
\hline
$\nu_\text{AP1}$ & 0.002 & - & $\nu_\text{DET}$ & 0.97 & 0.985  \\
$\mu_\text{AP1}$ & 520 ns & - & $\mu_\text{DET}$ & - & -1.8 ns \\
$\sigma_\text{AP1}$ & 90 ns  & - & $\sigma_\text{DET}$ & - & 5.1 \\
$\nu_\text{AP2}$ & 0.02  & - & $\nu_\text{DP}$ & 0.03 & 0.015 \\
$\mu_\text{AP2}$ & 1660 ns  & - & $\mu_\text{DP}$ & 58 & 48 ns \\
$\sigma_\text{AP2}$ & 680 ns  & - & $\sigma_\text{DP}$ & 5.3 & 10 ns \\
$\nu_\text{AP3}$ & 0.055 & 0.068 &   &  &   \\
$\mu_\text{AP3}$ & 6300 ns & 6703 ns & &  &  \\
$\sigma_\text{AP3}$ &1350 ns & 1229 ns &  &  &  \\
\hline
\end{tabular}
\end{center}
\label{tab:parinstrument}
\end{table}%

\begin{figure}[htbp]
\begin{center}
\includegraphics[width=\columnwidth]{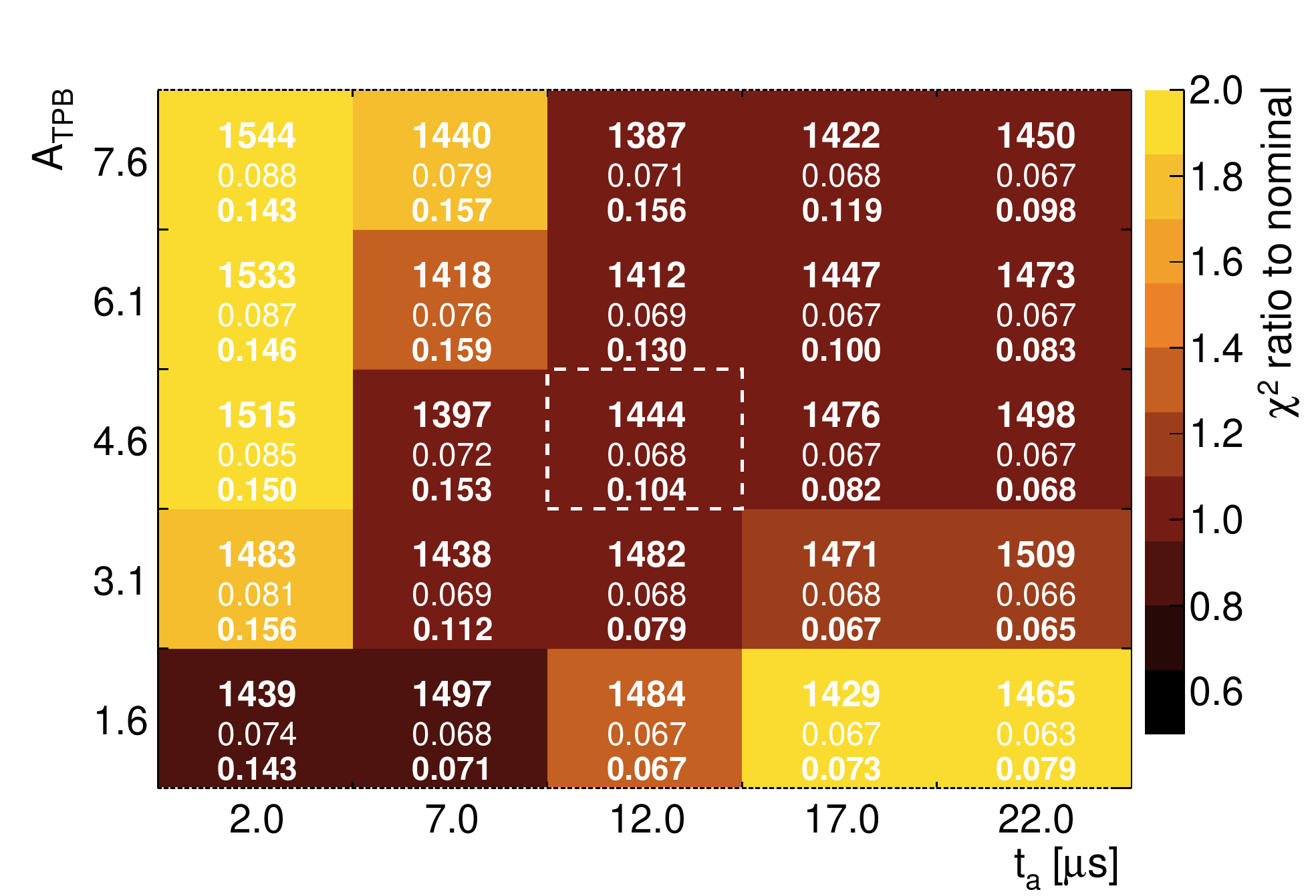}
\caption{Each box corresponds to a fit where $A_{TPB}$ and $t_a$ are fixed to the value indicated on the axis. The box in the very center (light dashed outline) corresponds to the nominal fit where $A_{TPB}$ and $t_a$ are fixed to the best fit values from \cite{Stanford:2018un}. The values on the axes span the range from $-2\sigma$ to $+2\sigma$ using the parameter uncertainties from \cite{Stanford:2018un}. A measure for the typical relative difference between model and data is shown on the color scale (see text for an explanation of how this is calculated). The fit-out values for $\tau_{t}$ (in units of \SI{}{\nano\second}), $\nu_\text{AP3}$, and $R_\text{TPB}$ are printed in each box, in that order from top to bottom.}
\label{fig:chi2map}
\end{center}
\end{figure}

The LAr triplet lifetime is strongly correlated in the fit with the TPB parameters $A_{TPB}$ and $t_a$, and with the afterpulsing probability in the 3rd afterpulsing region, $\nu_{AP3}$. To investigate how much effect the TPB parameters have on the LAr triplet lifetime, we varied $A_{TPB}$ and $t_a$ each within $\pm 2\sigma$ using the parameter uncertainties from \cite{Stanford:2018un}. For each combination of these parameter values, a fit was performed with all parameters fixed but for: $\tau_{t}$, $\nu_{AP3}$, $R_s$, $R_t$, $R_{TPB}$, and the overall normalization $I_0$. The resulting parameter values are shown in Fig.~\ref{fig:chi2map} on a grid with the test values of $t_a$ on the x-axis and the test values of $A_{TPB}$ on the y axis. The fit values for $\tau_{t}$, $\nu_{AP3}$, and $R_{TPB}$ are printed in each box, and the box is shaded by the ratio of the given fit's  $\chi^2$ to the value of  $\chi^2$ from the nominal fit. While in this case, the reduced $\chi^2$ parameter cannot be used to infer a p-value, the relative difference for different model parameters is still a useful quantity saying something about how close the model comes to the data. The box in the very center, outlined with a dashed line, is the nominal fit.

\begin{figure}[htbp]
\begin{center}
\includegraphics[width=\columnwidth]{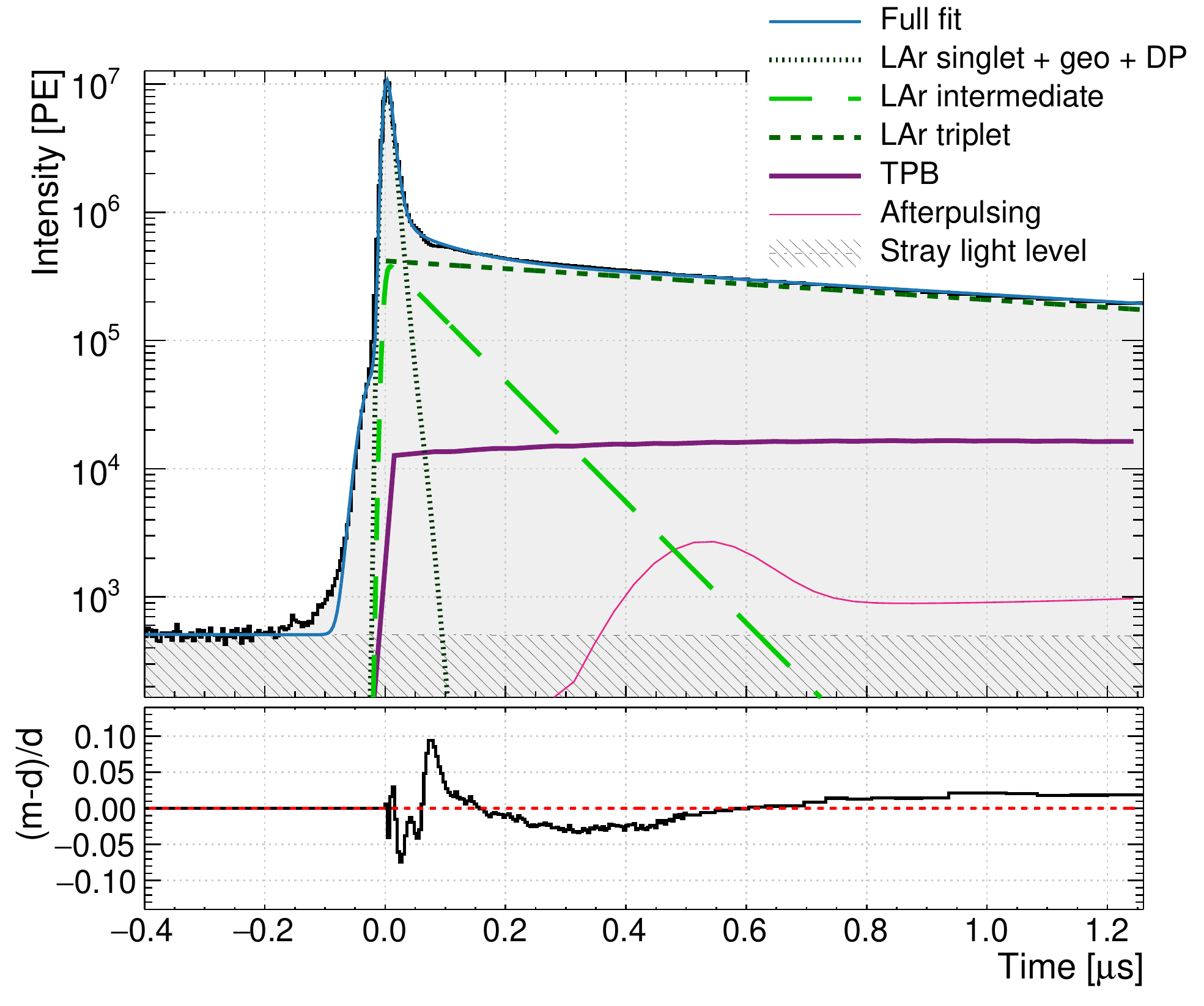}
\caption{The prompt region of the pulseshape is shown again but the shape of the LAr intermediate is described as a single exponential decay.}
\label{fig:APSshortfitExpIntermediate} \end{center}
\end{figure}

Fig.~\ref{fig:APSshortfitExpIntermediate} shows the fit with nominal parameters, but the shape of the LAr intermediate component is changed to a simple exponential decay. The ratio of $\chi^2$ between this fit and the nominal fit is 1.2, but reaches this level only if the late pulsing probability is allowed to vanish. The triplet lifetime in this fit is \SI{1435}{\nano\second}.

\section{Discussion}\label{sec:discussion}

The model described in Sect.~\ref{sec:fitmodel} fits the observed pulseshape with deviations between model and data of less than 11\% between \SIrange{0}{160}{\micro\second}. The most significant deviation occurs in the time range of \SIrange{15}{50}{\micro\second}. This time region is dominated by the delayed TPB emission, whose time-structure was described using the physics-based model and parameters from \cite{Stanford:2018un}. In \cite{Stanford:2018un}, the TPB model does not describe the TPB pulseshape between \SIrange{15}{50}{\micro\second} perfectly, either. Varying the model parameters within the uncertainties given, as shown in Fig.~\ref{fig:chi2map}, lead to a slight improvement in the fit for some combinations, since the fit can compensate by changing the values of the free parameters. An alternate model for the delayed TPB emission is proposed in \cite{2015PhRvC..91c5503S}. This effective model is based on a sum of exponential functions fit to the TPB emission pulseshape up to \SI{10}{\micro\second}, and fails to describe the pulseshapes discussed here for times $t \geq$~\SI{10}{\micro\second}. We note that the delayed TPB emission may be subject to quenching by electronegative impurities such as oxygen. Therefore, the time structure measured in two experiments with different impurity profiles may differ.

 The existence of delayed TPB emission means that each event contains light from previous events. For a \SI{10}{\micro\second} analysis window, approximately 3\% of the total light intensity is emitted after the nominal end of the event. To be sure to analyze only events free from light belonging to previous events, event-time cuts of more than \SI{200}{\micro\second} must be chosen. With a background rate of \SI{3300}{\hertz} due to \Nuc{Ar}{39} decay in DEAP-3600, such a long event time cut removes too much livetime to be viable. The delayed light from previous events appears as time-variable uncorrelated noise in the analysis, in addition to the constant dark rate. We determined this stray light level by studying the light intensity in the pre-event window of each event as a function of the time difference to the previous event. The level depends strongly on the total event rate in the detector, the energy spectrum of events in the detector, and on the pile-up cuts used. The time profile for stray light found using this method under-estimates the observed light level in the pulseshape at late times by approximately 10\%. This is likely due to subtle effects related to the trigger and data-quality cuts, since we compare the average pulse count very late in the pulseshape made from events selected by careful data-cleaning cuts to the average pulse count in the trace before the start of each event, with no control over what happened in the detector before the event. Particularly AP from light detected late in the previous event can increase the stray light rate as measured before the event peak.

The stray light component introduces subtle effects into the data. For example, given the dark noise rate of the PMTs, one would expect to measure on average 0.4~PE of uncorrelated noise in the \SI{10}{\micro\second} standard analysis window. Due to the long TPB decay component, the actual uncorrelated noise level is higher, and varies with the overall event rate and spectrum, in a way consistent with predictions of a toy Monte Carlo simulation. In regular physics data, on average \SI[separate-uncertainty=true]{1.3\pm0.1}{PE} of uncorrelated noise are measured per event. The uncertainty accounts for the slight mismatch between the stray light model and the pulseshape data. During detector calibrations with radioactive sources, the energy spectrum is changed and the rate of events is increased, such that the uncorrelated noise level is higher. For the calibration with the \Nuc{Na}{22} source for example, it comes to 2.6 PE. This leads to systematic differences at the percent level in energy calibrations done with different types of calibration sources. 

At low numbers of PE, PSD is also sensitive to the uncorrelated noise level. PSD in LAr is often based on the fraction of light detected in a prompt time window of $\mathcal{O}$(100ns) around the event peak (Fprompt). Consider an event at 80 PE total, 30\% of which occurs in the prompt window, so that the Fprompt parameter is $0.3$. With 0.4 additional PE, of which, due to the length of the prompt and late windows, 10\% occur in the prompt window and 90\% in the late window, the measured Fprompt is $0.299$\footnote{The effect of instrumental biases on the Fprompt parameter is discussed in \cite{d1psdpaper}}. With 1.3 additional PE, the measured Fprompt value is $0.295$ and with 2.6 additional PE, it is $0.290$. A 1\% to 3\% energy-dependent shift in the value of the PSD parameter can result in a noticeable systematic effect in background leakage predictions.

The overall structure of the pulseshape between approximately \SIrange{0.2}{10}{\micro\second} is well-described by the sum of the LAr triplet component, the TPB late time structure, and PMT AP. Periodic structures in the model-data comparison in the AP time regions are expected, since the AP time distribution has sub-features that the simple gaussian model from Eq.~\ref{eq:ap} does not capture. The 10\% discrepancy at \SI{12}{\micro\second} falls at the intersection of the AP and AP-of-AP regions and might relate to subtleties in the AP-of-AP mechanism that are not modelled here.

The lifetime of the LAr triplet state we measure here is \SI{1445}{\nano\second}. The statistical uncertainty is negligible, however there are large systematic uncertainties: the LAr triplet lifetime is correlated with the parameters of the delayed TPB component, so the result is sensitive to whether or not this component is included in the analysis, and to the assumed time structure.  As seen in Fig.~\ref{fig:chi2map}, the triplet lifetime varies between \SIrange{1387}{1544}{\nano\second} when varying the delayed TPB emission parameters within their uncertainties. Removing the delayed TPB component from the fit, we measure a triplet lifetime of \SI{1564}{\nano\second}. Literature values range from \SI{1300}{\nano\second} \cite{Hofmann:2013hf} to values near \SI{1600}{\nano\second} \cite{Hitachi:1983ja}\footnote{Several earlier measurements find smaller values near \SI{1000}{\nano\second} probably due to uncontrolled-for impurities in the LAr.}. The measurement in \cite{Hofmann:2013hf} was done without the use of a wavelength shifter, while all the measurements that find values of \SI{1500}{\nano\second} or more use TPB and assume that TPB re-emits all photons within a few nanoseconds. We also note that the LAr triplet lifetime one infers is strongly dependent on the level and kind of impurities in the LAr \cite{Amsler:2008jq,Acciarri:2010gm,Acciarri:2010jh,Peiffer:2008zz,Jones:2013bp}.

Near the event peak, instrumental effects compound such that the value of $\tau_p$ given in Tab.~\ref{tab:parphysics} must be understood as a combination of the LAr singlet decay, TPB prompt emission, and scattering effects in the detector.

We find that the model including a LAr intermediate component~(\cite{Hofmann:2013hf}) described in Eq.~\ref{eq:purelar} and surrounding text better describes our data than a simple exponential decay model. 
The hypothesis of delayed recombination could be tested by studying the pulseshape in a detector where an electric drift field can be applied. For a field high enough to drift all ionization electrons away from the interaction region, the intermediate component should disappear altogether. 

If the hypothesis about delayed recombination is correct, the shape and intensity of the intermediate component should change with linear energy transfer. This would in principle offer another PSD-based handle on separating, for example, nuclear recoils from electron-recoil backgrounds. However, since this component does not dominate the pulseshape at any time, and only plays a role in a small time window, in practice, no PSD power improvement due to it should be expected. However, it should be taken into account when optimizing the length of the prompt window for Fprompt-like PSD parameters.

\section{Conclusion}\label{sec:conclusion}
We present a complete model for the overall features of the pulseshape observed in a large LAr-based particle detector using TPB for wavelength shifting and PMTs for photon detection. The model accounts for the LAr intermediate component and delayed TPB emission. The existence of delayed TPB emission has been proposed from dedicated small-scale setups and is verified and measured here for the first time in a large detector. It has consequences for the interpretation of energy calibrations, and for particle identification through PSD. It also influences practical detector operation and design decisions, such as the length of the event windows and the pile-up rate, which in part determines the ultimate size limit on a detector. It must therefore be taken into account in interpreting results from LAr-based detectors, and in planning for future detectors, all of which currently use, or plan to use, TPB for wavelength shifting.

The model can also be used to understand detector behaviour by enabling a correct implementation of the signal shape in detector Monte Carlo simulation. The fits to the pulseshapes can be used to monitor instrumental effects, such as afterpulsing in PMTs, with fine time-resolution and without the need for dedicated calibration data, due to the large rate of \Nuc{Ar}{39} $\beta$-decays available for analysis.

\section*{Acknowledgements}
We thank the Natural Sciences
  and Engineering Research Council of Canada, the Canadian Foundation
  for Innovation (CFI), the Ontario Ministry of Research and Innovation (MRI), and
  Alberta Advanced Education and Technology (ASRIP), Queen's
  University, the University of Alberta, Carleton University, the
  Canada First Research Excellence Fund, the Arthur B.~McDonald
  Canadian Astroparticle Research Institute, DGAPA-UNAM (PAPIIT No. IA100118 and IN108020) and Consejo Nacional de Ciencia y Tecnolog{\'i}a (CONACyT, Mexico, Grants No. 252167 and A1-S-8960), the European
  Research Council Project (ERC StG 279980), the UK Science and
  Technology Facilities Council (STFC ST/K002570/1 and ST/R002908/1), the Russian Science Foundation (Grant No 16-12-10369), the Leverhulme Trust (ECF-20130496), the Spanish Ministry of Science, 
  Innovation and Universities (FPA2017-82647-P grant and MDM-2015-0509), and the International Research Agenda Programme AstroCeNT (MAB/2018/7) funded by the Foundation for Polish Science (FNP) from the European Regional Development Fund.
  Studentship support from the Rutherford
  Appleton Laboratory Particle Physics Division, STFC and SEPNet
  PhD is acknowledged.  We would like to thank SNOLAB and its
  staff for support through underground space, logistical, and
  technical services. SNOLAB operations are supported by the CFI and 
 Province of Ontario MRI, with underground access provided by Vale at
  the Creighton mine site. We thank Vale for support in shipping the
  acrylic vessel underground.  We gratefully acknowledge the support
  of Compute Canada, Calcul Qu\'ebec, the Center for Advanced
  Computing at Queen's University, and the Computation Center for Particle 
  and Astrophysics (C2PAP) at the Leibniz Supercomputer Center (LRZ) for
  providing the computing resources required to undertake this work.

\bibliography{apspaper_bib_complete}

\begin{thebibliography}{10}

\bibitem{Fiorillo:2006bt}
G~Fiorillo.
\newblock
  \href{http://adsabs.harvard.edu/cgi-bin/nph-data_query?bibcode=2006NuPhS.150..372F&link_type=ABSTRACT}{The
  liquid argon technology for neutrino and astroparticle detectors}.
\newblock In {\em Nuc. Phys. B Proc. Sup.}, Dipartimento di Scienze Fisiche
  Universit{\`a} di Napoli "Federico II" and INFN Sezione di Napoli, Italy,
  2006, pages 372--376.

\bibitem{Gary:2013bj}
C~Gary, S~Kane, M~I Firestone, et~al.
\newblock \href{https://doi.org/10.1063/1.4802417}{{Large area liquid argon
  detectors for interrogation systems}}.
\newblock In {\em Application of accelerators in research and industry: 22.
  International Conference}, AIP, 2013, pages 698--703.

\bibitem{Agostini2015jf}
M~Agostini, M~Barnab{\'e}-Heider, D~Budj{\'a}{\v s}, et~al.
\newblock
  \href{https://link.springer.com/article/10.1140/epjc/s10052-015-3681-5}{{LArGe:
  active background suppression using argon scintillation for the GERDA $0\nu
  \beta \beta $-experiment}}.
\newblock {\em EPJ C}, 75\penalty0 (10):\penalty0 506, 2015,
  \href{http://arxiv.org/abs/1501.05762}{{\ttfamily arXiv:1501.05762}}.

\bibitem{Meyers:2015dc}
{\bfseries DarkSide-50} Collaboration, P~Agnes, D~Alton, K~Arisaka, et~al.
\newblock
  \href{http://pubget.com/site/paper/1b2415f8-d585-42a5-a33a-c8381b41b0b3?institution=}{{DarkSide-50:
  A WIMP Search with a Two-phase Argon TPC}}.
\newblock {\em Physics Procedia}, 61:\penalty0 124--129, 2015.

\bibitem{Aalseth:2017um}
{\bfseries DarkSide-20k} Collaboration, C.~E. Aalseth, F~Acerbi, P~Agnes,
  et~al.
\newblock \href{http://arxiv.org/abs/1707.08145}{{DarkSide-20k: A 20 Tonne
  Two-Phase LAr TPC for Direct Dark Matter Detection at LNGS}}.
\newblock {\em EPJ Plus}, 133:\penalty0 131, 2017,
  \href{http://arxiv.org/abs/1707.08145}{{\ttfamily arXiv:1707.08145}}.

\bibitem{gerda}
{\bfseries {GERDA}} Collaboration, M~Agostini, Bakalyarov A.M., M~Balata,
  et~al.
\newblock \href{https://doi.org/{10.1140/epjc/s10052-018-5812-2}}{{Upgrade for
  Phase II of the Gerda experiment}}.
\newblock {\em EPJ C}, 78\penalty0 (5), 2017,
  \href{http://arxiv.org/abs/1711.01452}{{\ttfamily arXiv:1711.01452}}.

\bibitem{legend}
{\bfseries {LEGEND}} Collaboration, N~Abgrall, A~Abramov, N~Abrosimov, et~al.
\newblock \href{https://doi.org/{10.1063/1.5007652}}{{The Large Enriched
  Germanium Experiment for Neutrinoless Double Beta Decay (LEGEND)}}.
\newblock {\em AIP Conf.Proc.}, 1894\penalty0 (1), 2017,
  \href{http://arxiv.org/abs/1709.01980}{{\ttfamily arXiv:1709.01980}}.

\bibitem{Collaboration:2016ty}
{\bfseries DUNE} Collaboration, R~Acciarri, MA~Acero, M~Adamowski, et~al.
\newblock \href{http://arxiv.org/abs/1601.05471}{{Long-Baseline Neutrino
  Facility (LBNF) and Deep Underground Neutrino Experiment (DUNE) Conceptual
  Design Report Volume 1}}.
\newblock 2016, \href{http://arxiv.org/abs/1601.05471}{{\ttfamily
  arXiv:1601.05471}}.

\bibitem{Akimov:2018ghi}
D~Akimov, J~B Albert, P~An, et~al.
\newblock \href{http://arxiv.org/abs/1803.09183v2}{{COHERENT 2018 at the
  Spallation Neutron Source}}.
\newblock 2018, \href{http://arxiv.org/abs/1803.09183v2}{{\ttfamily
  arXiv:1803.09183v2}}.

\bibitem{Ajaj:2019wi}
{\bfseries DEAP} Collaboration, R~Ajaj, P~A Amaudruz, G~R Araujo, et~al.
\newblock \href{https://doi.org/10.1103/PhysRevD.100.022004}{{Search for dark
  matter with a 231-day exposure of liquid argon using DEAP-3600 at SNOLAB}}.
\newblock {\em Phys. Rev. D}, 100:\penalty0 022004, 2019,
  \href{http://arxiv.org/abs/1902.04048}{{\ttfamily arXiv:1902.04048}}.

\bibitem{d1psdpaper}
{\bfseries {DEAP}} Collaboration, M~G Boulay, B~Cai, M~Chen, et~al.
\newblock
  \href{https://linkinghub.elsevier.com/retrieve/pii/S0927650516301232}{{Measurement
  of the scintillation time spectra and pulse-shape discrimination of
  low-energy $\beta$ and nuclear recoils in liquid argon with DEAP-1}}.
\newblock {\em Astroparticle Physics}, 85:\penalty0 1--23, 2016,
  \href{http://arxiv.org/abs/0904.2930}{{\ttfamily arXiv:0904.2930}}.

\bibitem{Lippincott:2008uaa}
W~H Lippincott, K.~J. Coakley, D~Gastler, et~al.
\newblock \href{https://doi.org/10.1103/PhysRevC.78.035801
  10.1103/PhysRevC.81.039901}{{Scintillation time dependence and pulse shape
  discrimination in liquid argon}}.
\newblock {\em Phys. Rev. C}, 78:\penalty0 035801, 2008,
  \href{http://arxiv.org/abs/0801.1531}{{\ttfamily arXiv:0801.1531}}.

\bibitem{Carvalho:1979tm}
MJ~Carvalho and G~Klein.
\newblock \href{https://doi.org/10.1016/0022-2313(79)90167-4}{{Luminescence
  decay in condensed argon under high energy excitation}}.
\newblock {\em Journal of Luminescence}, 18/19:\penalty0 487--490, 1979.

\bibitem{Kubota:1978kh}
S~Kubota, M~Hishida, and J~Raun.
\newblock \href{https://doi.org/10.1088/0022-3719/11/12/024}{{Evidence for a
  triplet state of the self-trapped exciton states in liquid argon, krypton and
  xenon}}.
\newblock {\em J. Phys. C}, 11:\penalty0 2645, 1978.

\bibitem{Morikawa:1989gv}
E~Morikawa, R~Reininger, P~G{\"u}rtler, V~Saile, and P~Laporte.
\newblock \href{https://doi.org/10.1063/1.457108}{{Argon, krypton, and xenon
  excimer luminescence: From the dilute gas to the condensed phase}}.
\newblock {\em J. Chem. Phys.}, 91\penalty0 (3):\penalty0 1469, 1989.

\bibitem{Hitachi:1983ja}
Akira Hitachi, Tan Takahashi, Nobutaka Funayama, et~al.
\newblock \href{https://doi.org/10.1103/PhysRevB.27.5279}{{Effect of ionization
  density on the time dependence of luminescence from liquid argon and xenon}}.
\newblock {\em Phys. Rev. B}, 27:\penalty0 5279, 1983.

\bibitem{Peiffer:2008zz}
P~Peiffer, T~Pollmann, S~Sch{\"o}nert, A.~Smolnikov, and S.~Vasiliev.
\newblock
  \href{https://iopscience.iop.org/article/10.1088/1748-0221/3/08/P08007}{{Pulse
  shape analysis of scintillation signals from pure and xenon-doped liquid
  argon for radioactive background identification}}.
\newblock {\em JINST}, 3:\penalty0 P08007, 2008.

\bibitem{Acciarri:2010gm}
R~Acciarri, M~Antonello, B~Baibussinov, et~al.
\newblock \href{https://doi.org/10.1088/1748-0221/5/06/P06003}{{Effects of
  Nitrogen contamination in liquid Argon}}.
\newblock {\em JINST}, 5\penalty0 (06):\penalty0 P06003--P06003, 2010,
  \href{http://arxiv.org/abs/0804.1217}{{\ttfamily arXiv:0804.1217}}.

\bibitem{Hofmann:2013hf}
M~Hofmann, T~Dandl, T~Heindl, et~al.
\newblock \href{https://doi.org/10.1140/epjc/s10052-013-2618-0}{{Ion-beam
  excitation of liquid argon}}.
\newblock {\em EPJ C}, 73\penalty0 (10):\penalty0 2618, 2013,
  \href{http://arxiv.org/abs/1511.07721}{{\ttfamily arXiv:1511.07721}}.

\bibitem{2015PhRvC..91c5503S}
E~Segreto.
\newblock
  \href{http://adsabs.harvard.edu/cgi-bin/nph-data_query?bibcode=2015PhRvC..91c5503S&link_type=ABSTRACT}{{Evidence
  of delayed light emission of tetraphenyl-butadiene excited by liquid-argon
  scintillation light}}.
\newblock {\em Phys. Rev. C}, 91\penalty0 (3):\penalty0 035503, 2015,
  \href{http://arxiv.org/abs/1411.4524}{{\ttfamily arXiv:1411.4524}}.

\bibitem{Burton:1973up}
B~A Powell and W~M Burton.
\newblock \href{https://doi.org/10.1364/AO.12.000087}{{Fluorescence of
  Tetraphenyl-Butadiene in the Vacuum Ultraviolet}}.
\newblock {\em Applied Optics}, 12\penalty0 (1):\penalty0 87--89, 1973.

\bibitem{Davies:1996vd}
G~J Davies, C~H Lally, W~G Jones, and N~J~T Smith.
\newblock
  \href{http://adsabs.harvard.edu/cgi-bin/nph-data_query?bibcode=1996NIMPB.117..421D&link_type=ABSTRACT}{{UV
  quantum efficiencies of organic fluors}}.
\newblock {\em NIM B}, 117, 1996.

\bibitem{TPBpaper}
T~Pollmann, M~Boulay, and M.~Kuzniak.
\newblock \href{https://doi.org/10.1016/j.nima.2011.01.045}{{Scintillation of
  thin tetraphenyl butadiene films under alpha particle excitation}}.
\newblock {\em NIM A}, 635\penalty0 (1):\penalty0 127--130, 2011,
  \href{http://arxiv.org/abs/1011.1012}{{\ttfamily arXiv:1011.1012}}.

\bibitem{Veloce:2015slj}
L~M Veloce, M.~Kuzniak, P~C F~Di Stefano, et~al.
\newblock \href{https://doi.org/10.1088/1748-0221/11/06/P06003}{{Temperature
  dependence of alpha-induced scintillation in the
  1,1,4,4-tetraphenyl-1,3-butadiene wavelength shifter}}.
\newblock {\em JINST}, 11\penalty0 (06):\penalty0 P06003--P06003, 2016,
  \href{http://arxiv.org/abs/1511.08424}{{\ttfamily arXiv:1511.08424}}.

\bibitem{Stanford:2018un}
C~Stanford, S~Westerdale, J~Xu, and F~Calaprice.
\newblock \href{https://doi.org/10.1103/PhysRevD.98.062002}{{Surface background
  suppression in liquid argon dark matter detectors using a newly discovered
  time component of tetraphenyl-butadiene scintillation }}.
\newblock {\em Phys. Rev. D}, 98\penalty0 (ins-det), 2018,
  \href{http://arxiv.org/abs/1804.06895v1}{{\ttfamily arXiv:1804.06895v1}}.

\bibitem{Asaadi:2018vz}
J~Asaadi, B~J~P Jones, A~Tripathi, et~al.
\newblock \href{https://doi.org/10.1088/1748-0221/14/02/P02021}{Tetraphenyl
  butadiene emanation and bulk fluorescence from wavelength shifting coatings
  in liquid argon}.
\newblock {\em JINST}, 14\penalty0 (ins-det), 2019,
  \href{http://arxiv.org/abs/1804.00011}{{\ttfamily arXiv:1804.00011}}.

\bibitem{detectorpaper}
{\bfseries {DEAP}} Collaboration, P~A Amaudruz, M~Baldwin, M~Batygov, et~al.
\newblock
  \href{https://linkinghub.elsevier.com/retrieve/pii/S0927650518300914}{{Design
  and construction of the DEAP-3600 dark matter detector}}.
\newblock {\em Astroparticle Physics}, 108:\penalty0 1--23, 2019,
  \href{http://arxiv.org/abs/1712.01982}{{\ttfamily arXiv:1712.01982}}.

\bibitem{Broerman:2017hf}
B~Broerman, M~G Boulay, B~Cai, et~al.
\newblock \href{https://doi.org/10.1088/1748-0221/12/04/p04017}{{Application of
  the {TPB} Wavelength Shifter to the DEAP-3600 Spherical Acrylic Vessel Inner
  Surface }}.
\newblock {\em Journal of Instrumentation}, 12:\penalty0 04017, 2017,
  \href{http://arxiv.org/abs/1704.01882}{{\ttfamily arXiv:1704.01882}}.

\bibitem{finemet}
Hitachi.
\newblock
  \href{{https://www.hitachi-metals.co.jp/e/products/elec/tel/p02\_21.html}}{{Nanocrystalline
  soft magnetic material, FINEMET}}.
\newblock {\em
  {https://www.hitachi-metals.co.jp/e/products/elec/tel/p02\_21.html}}.

\bibitem{Ajaj:2019to}
{\bfseries {DEAP}} Collaboration, R~Ajaj, G~R Araujo, M~Batygov, et~al.
\newblock \href{http://arxiv.org/abs/1905.05811}{{Electromagnetic Backgrounds
  and Potassium-42 Activity in the DEAP-3600 Dark Matter Detector}}.
\newblock {\em Phys. Rev. D}, 100:\penalty0 072009, 2019,
  \href{http://arxiv.org/abs/1905.05811}{{\ttfamily arxiv:1905.05811}}.

\bibitem{Hitachi:1992ve}
A~Hitachi, T~Doke, and A~Mozumder.
\newblock
  \href{http://link.aps.org/doi/10.1103/PhysRevB.46.11463}{{Luminescence
  quenching in liquid argon under charged-particle impact: Relative
  scintillation yield at different linear energy transfers}}.
\newblock {\em Phys. Rev. B}, 46\penalty0 (18), 1992.

\bibitem{Ribitzki:1994de}
G~Ribitzki, A~Ulrich, B~Busch, et~al.
\newblock {Electron densities and temperatures in a xenon afterglow with
  heavy-ion excitation}.
\newblock {\em Phys. Rev. E}, 50\penalty0 (5):\penalty0 3973--3979, 1994.

\bibitem{Kubota:1979gk}
Shinzou Kubota, Masahiko Hishida, Masayo Suzuki, and Jian-Zhi Ruan.
\newblock {Dynamical behavior of free electrons in the recombination process in
  liquid argon, krypton, and xenon}.
\newblock {\em Phys. Rev. B}, 20:\penalty0 3486, 1979.

\bibitem{hofmann}
Martin Hofmann.
\newblock {\em {Liquid Scintillators and Liquefied Rare Gases for Particle
  Detectors}}.
\newblock PhD thesis, Technical Unviersity of Munich, 2012.

\bibitem{pmtpaper}
{\bfseries DEAP} Collaboration, P~A Amaudruz, M~Batygov, B~Beltran, et~al.
\newblock \href{https://doi.org/10.1016/j.nima.2018.12.058}{{In-situ
  characterization of the Hamamatsu R5912-HQE photomultiplier tubes used in the
  DEAP-3600 experiment}}.
\newblock {\em NIM A}, 922:\penalty0 373--384, 2019,
  \href{http://arxiv.org/abs/1705.10183}{{\ttfamily arXiv:1705.10183}}.

\bibitem{Amsler:2008jq}
C~Amsler, V~Boccone, A~B{\"u}chler, et~al.
\newblock \href{https://doi.org/10.1088/1748-0221/3/02/P02001}{{Luminescence
  quenching of the triplet excimer state by air traces in gaseous argon}}.
\newblock {\em JINST}, 3:\penalty0 2001, 2008,
  \href{http://arxiv.org/abs/0708.2621}{{\ttfamily arXiv:0708.2621}}.

\bibitem{Acciarri:2010jh}
R~Acciarri, M~Antonello, B~Baibussinov, et~al.
\newblock \href{https://doi.org/10.1088/1748-0221/5/05/P05003}{{Oxygen
  contamination in liquid Argon: combined effects on ionization electron charge
  and scintillation light}}.
\newblock {\em JINST}, 5\penalty0 (05):\penalty0 P05003--P05003, 2010,
  \href{http://arxiv.org/abs/0804.1222}{{\ttfamily arXiv:0804.1222}}.

\bibitem{Jones:2013bp}
B~J~P Jones, T~Alexander, H~O Back, et~al.
\newblock \href{https://doi.org/10.1088/1748-0221/8/12/P12015}{{The effects of
  dissolved methane upon liquid argon scintillation light}}.
\newblock {\em JINST}, 8\penalty0 (12):\penalty0 P12015--P12015, 2013,
  \href{http://arxiv.org/abs/1308.3658}{{\ttfamily arXiv:1308.3658}}.

\end{thebibliography}
\bibliographystyle{tp_unsrt_doi}

\end{document}